\documentclass[a4paper, amsfonts, amssymb, amsmath, reprint, prl, aps,
			   showkeys, nofootinbib, superscriptaddress, twoside]{revtex4-2}
\usepackage{amsthm}
\usepackage{mathtools}
\usepackage{physics}
\usepackage{xcolor}
\usepackage{graphicx}
\usepackage[left=23mm,right=13mm,top=35mm,columnsep=15pt]{geometry} 
\usepackage{adjustbox}
\usepackage{placeins}
\usepackage[T1]{fontenc}
\usepackage{lipsum}
\usepackage[english]{babel}
\usepackage[utf8]{inputenc}
\usepackage{csquotes}
\usepackage[mathscr]{eucal} 
\usepackage{bibunits} 
\usepackage[pdftex, pdftitle={Article}, pdfauthor={Author}]{hyperref} 
\usepackage[capitalize]{cleveref}
\usepackage{multirow}
\newcommand{\zeroL}{\ket{0_{\mathrm{L}}}}
\newcommand{\oneL}{\ket{1_{\mathrm{L}}}}
\newcommand{\plusL}{\ket{+_{\mathrm{L}}}}
\newcommand{\minusL}{\ket{-_{\mathrm{L}}}}
\newcommand{\iplusL}{\ket{+i_{\mathrm{L}}}}
\newcommand{\iminusL}{\ket{-i_{\mathrm{L}}}}
\newcommand{\ZL}{Z_{\mathrm{L}}}
\newcommand{\XL}{X_{\mathrm{L}}}
\newcommand{\YL}{Y_{\mathrm{L}}}
\newcommand{\avgZL}{\langle Z_{\mathrm{L}}\rangle}
\newcommand{\avgXL}{\langle X_{\mathrm{L}}\rangle}
\newcommand{\avgYL}{\langle Y_{\mathrm{L}}\rangle}
\newcommand{\TL}{T_{\mathrm{L}}}
\newcommand{\ZD}[1]{Z_{\mathrm{D} #1 }}
\newcommand{\XD}[1]{X_{\mathrm{D} #1 }}
\newcommand{\YD}[1]{Y_{\mathrm{D} #1 }}
\newcommand{\Done}{\mathrm{D}_{1}}
\newcommand{\Dtwo}{\mathrm{D}_{2}}
\newcommand{\Dthree}{\mathrm{D}_{3}}
\newcommand{\Dfour}{\mathrm{D}_{4}}
\newcommand{\Aone}{\mathrm{A}_{1}}
\newcommand{\Atwo}{\mathrm{A}_{2}}
\newcommand{\Athree}{\mathrm{A}_{3}}
\newcommand{\Zone}{\mathrm{A}_{1}}
\newcommand{\Ztwo}{\mathrm{A}_{3}}
\newcommand{\X}{\mathrm{A}_{2}}

\newcommand{\ns}{\mathrm{ns}}
\newcommand{\us}{\mathrm{\mu s}}

\newcommand{\Tone}{T_\mathrm{1}}
\newcommand{\Ttwoecho}{T_\mathrm{2}}
\newcommand{\Ttwostar}{T^*_\mathrm{2}}
\newcommand{\Tphi}{T_{\mathrm{\phi}}}
\newcommand{\R}{\mathcal{R}}
\newcommand{\FidfourQ}{F_{\mathrm{4Q}}}
\newcommand{\FidL}{F_{\mathrm{L}}}
\newcommand{\FidLGate}{F_\mathrm{L}^\mathrm{G}}
\newcommand{\FidLReadout}{F_\mathrm{L}^\mathrm{R}}
\newcommand{\Gpip}{\gamma_\mathrm{pip}}
\newcommand{\Gpar}{\gamma_\mathrm{par}}
\newcommand{\leakrate}{L_{1}}

\newcommand{\prob}{P}
\newcommand{\qubit}{\mathrm{Q}}
\newcommand{\anc}{\mathrm{A}}
\newcommand{\phivec}{\vec{\phi}} 

\begin{document}
\title{Logical-qubit operations in an error-detecting surface code}
\author{J.~F.~Marques}
\affiliation{QuTech, Delft University of Technology, P.O. Box 5046, 2600 GA Delft, The Netherlands}
\affiliation{Kavli Institute of Nanoscience, Delft University of Technology, P.O. Box 5046, 2600 GA Delft, The Netherlands}

\author{B.~M.~Varbanov}
\affiliation{QuTech, Delft University of Technology, P.O. Box 5046, 2600 GA Delft, The Netherlands}

\author{M.~S.~Moreira}
\affiliation{QuTech, Delft University of Technology, P.O. Box 5046, 2600 GA Delft, The Netherlands}
\affiliation{Kavli Institute of Nanoscience, Delft University of Technology, P.O. Box 5046, 2600 GA Delft, The Netherlands}

\author{H.~Ali}
\affiliation{QuTech, Delft University of Technology, P.O. Box 5046, 2600 GA Delft, The Netherlands}
\affiliation{Kavli Institute of Nanoscience, Delft University of Technology, P.O. Box 5046, 2600 GA Delft, The Netherlands}

\author{N.~Muthusubramanian}
\affiliation{QuTech, Delft University of Technology, P.O. Box 5046, 2600 GA Delft, The Netherlands}
\affiliation{Kavli Institute of Nanoscience, Delft University of Technology, P.O. Box 5046, 2600 GA Delft, The Netherlands}

\author{C.~Zachariadis}
\affiliation{QuTech, Delft University of Technology, P.O. Box 5046, 2600 GA Delft, The Netherlands}
\affiliation{Kavli Institute of Nanoscience, Delft University of Technology, P.O. Box 5046, 2600 GA Delft, The Netherlands}

\author{F.~Battistel}
\affiliation{QuTech, Delft University of Technology, P.O. Box 5046, 2600 GA Delft, The Netherlands}

\author{M.~Beekman}
\affiliation{QuTech, Delft University of Technology, P.O. Box 5046, 2600 GA Delft, The Netherlands}
\affiliation{Netherlands Organisation for Applied Scientific Research (TNO), P.O. Box 96864, 2509 JG The Hague, The Netherlands}

\author{N.~Haider}
\affiliation{QuTech, Delft University of Technology, P.O. Box 5046, 2600 GA Delft, The Netherlands}
\affiliation{Netherlands Organisation for Applied Scientific Research (TNO), P.O. Box 96864, 2509 JG The Hague, The Netherlands}

\author{W.~Vlothuizen}
\affiliation{QuTech, Delft University of Technology, P.O. Box 5046, 2600 GA Delft, The Netherlands}
\affiliation{Netherlands Organisation for Applied Scientific Research (TNO), P.O. Box 96864, 2509 JG The Hague, The Netherlands}

\author{A.~Bruno}
\affiliation{QuTech, Delft University of Technology, P.O. Box 5046, 2600 GA Delft, The Netherlands}
\affiliation{Kavli Institute of Nanoscience, Delft University of Technology, P.O. Box 5046, 2600 GA Delft, The Netherlands}

\author{B.~M.~Terhal}
\affiliation{QuTech, Delft University of Technology, P.O. Box 5046, 2600 GA Delft, The Netherlands}
\affiliation{JARA Institute for Quantum Information, Forschungszentrum Juelich, D-52425 Juelich, Germany}

\author{L.~DiCarlo}
\affiliation{QuTech, Delft University of Technology, P.O. Box 5046, 2600 GA Delft, The Netherlands}
\affiliation{Kavli Institute of Nanoscience, Delft University of Technology, P.O. Box 5046, 2600 GA Delft, The Netherlands}

\date{\today}
\begin{abstract}
We realize a suite of logical operations on a distance-two logical qubit stabilized using repeated error detection cycles. Logical operations include initialization into arbitrary states, measurement in the cardinal bases of the Bloch sphere, and a universal set of single-qubit gates. For each type of operation, we observe higher performance for fault-tolerant variants over non-fault-tolerant variants, and quantify the difference through detailed characterization. In particular, we demonstrate process tomography of logical gates, using the notion of a logical Pauli transfer matrix. This integration of high-fidelity logical operations with a scalable scheme for repeated stabilization is a milestone on the road to quantum error correction with higher-distance superconducting surface codes.
\end{abstract}
\maketitle

\begin{bibunit}[apsrev4-2]
\section{Introduction} \label{sec:introduction}
    Two key capabilities will distinguish an error-corrected quantum computer from present-day noisy intermediate-scale quantum (NISQ) processors~\cite{Preskill18}. 
    First, it will initialize, transform, and measure quantum information encoded in logical qubits rather than physical qubits. A logical qubit is a highly entangled two-dimensional subspace in the larger Hilbert space of many more physical qubits. Second, it will use repetitive quantum parity checks to discretize, signal, and (with aid of a decoder) correct errors occurring in the constituent physical qubits without destroying the encoded information~\cite{Terhal15}. Provided the incidence of physical errors is below a code-specific threshold and the quantum circuits for logical operations and stabilization are fault-tolerant, the logical error rate can be exponentially suppressed by increasing the distance (redundancy) of the quantum error correction (QEC) code employed~\cite{Martinis15}. At present, the exponential suppression for specific physical qubit errors (bit-flip or phase-flip) has been experimentaly demonstrated~\cite{Kelly15, Chen21} for repetition codes~\cite{Riste15, Cramer16, Riste20}.
    
    Leading experimental quantum platforms have taken key steps towards implementing QEC codes protecting logical qubits from general physical qubit errors. In particular, trapped-ion systems have demonstrated logical-level initialization, gates and measurements for single logical qubits in the Calderbank-Shor-Steane~\cite{Nigg14} and Bacon-Shor~\cite{Egan20} codes. Most recently, entangling operations between two logical qubits have been demonstrated in the surface code using lattice surgery~\cite{Erhard21}. However, except for smaller-scale experiments using two ion species~\cite{Negnevitsky18}, trapped-ion experiments in QEC have so far been limited to a single round of stabilization.
    
    In parallel, taking advantage of highly-non-demolition measurement in circuit quantum electrodynamics~\cite{Blais04}, superconducting circuits have taken key strides in repetitive stabilization of two-qubit entanglement~\cite{Andersen19, Bultink20} and logical qubits.
    Quantum memories based on 3D-cavity logical qubits in cat~\cite{Ofek16, Hu19} and  Gottesman-Kitaev-Preskill~\cite{Campagne-Ibarcq20} codes have crossed the memory break-even point.  Meanwhile, monolithic architectures have focused on logical qubit stabilization in a surface code realized with a 2D lattice of transmon qubits. Currently, the surface code~\cite{Fowler12} is the most attractive QEC code for solid-state implementation owing to its practical nearest-neighbor connectivity requirement and high error threshold. Recent experiments~\cite{Andersen20, Chen21} have demonstrated repetitive stabilization by post-selection in a surface code which, owing to its small size, is capable of quantum error detection but not correction.
    
    We demonstrate a complete suite of logical-qubit operations for this small (distance-2) surface code while preserving multi-round stabilization. Our logical operations span initialization anywhere on the logical Bloch sphere, measurement in all cardinal bases, and a universal set of single-logical-qubit gates. For each type of operation, we quantify the increased performance of fault-tolerant variants over non-fault-tolerant ones. We introduce the notion of a logical Pauli transfer matrix to describe a logical gate, analogous to the procedure commonly used to describe gates on physical qubits~\cite{Chow12}. Finally, we compare the performance of two scalable, fault-tolerant stabilizer measurement schemes compatible with our quantum hardware architecture~\cite{Versluis17}.

\begin{figure}
		    \centering
		    \includegraphics[width=0.487\textwidth]{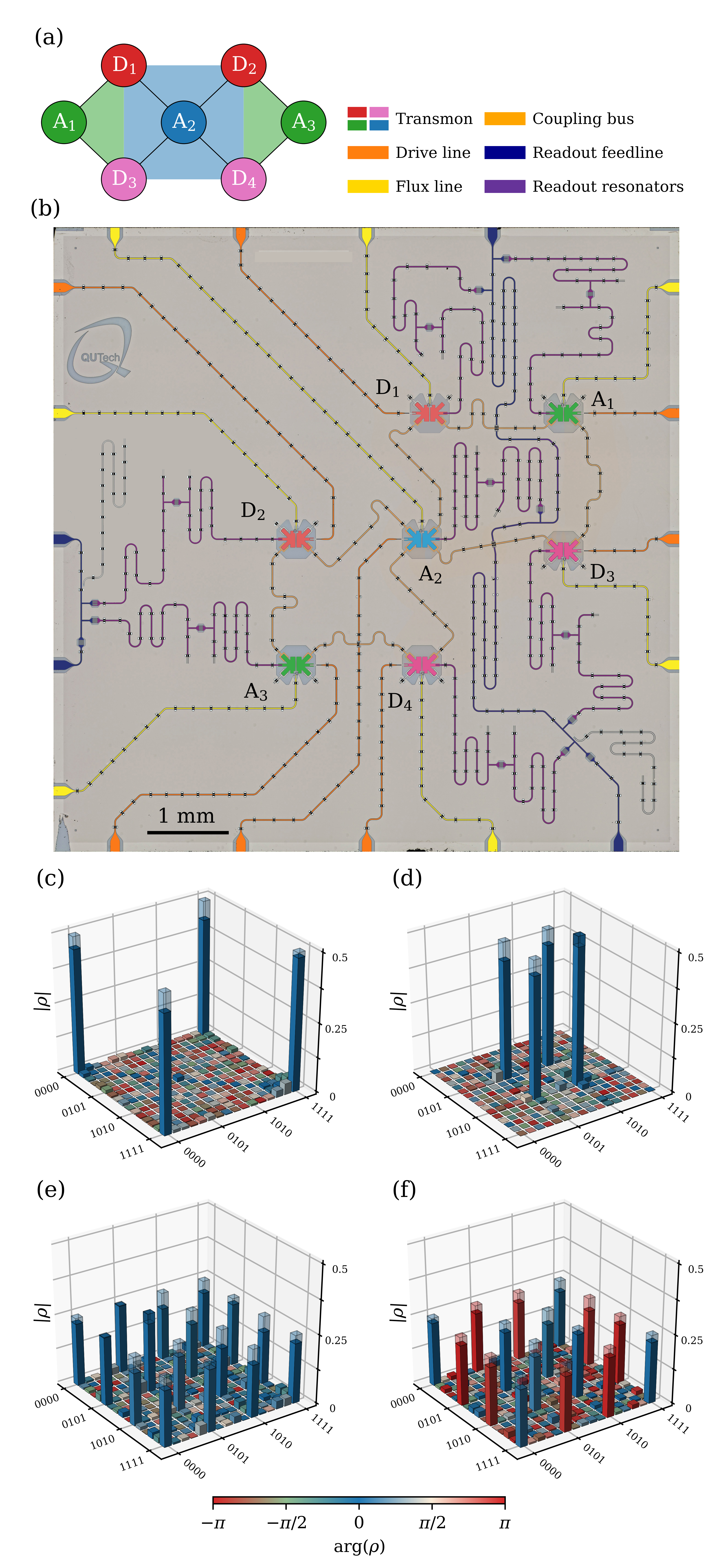}
		    \caption{\textbf{Surface-7 quantum processor and initialization of logical cardinal states.}
		    (a) Distance-two surface code. (b) Optical image of the quantum hardware with added false-color to emphasize different circuit elements.
		    (c-f) Estimated physical density matrices, $\rho$, after targeting the preparation of the logical cardinal states $\zeroL$ (c), $\oneL$ (d), $\plusL$ (e) and $\minusL$ (f). Each state is measured after preparing the data qubits in $\ket{0000}$, $\ket{1010}$, $\ket{+\!+\!++}$ and $\ket{+\!+\!--}$, respectively. The ideal target state density matrix is shown in the shaded wireframe.}
		    \label{fig:state_tomo}
		\end{figure}

	The distance-2 surface code (Fig.~\ref{fig:state_tomo}a) uses four data qubits ($\Done$ through $\Dfour$) to encode one logical qubit, whose two-dimensional codespace is the even-parity (i.e., eigenvalue +1) subspace of the stabilizer set
	\begin{equation}
	\mathscr{S} = \{\ZD{1}\ZD{3}, \XD{1}\XD{2}\XD{3}\XD{4}, \ZD{2}\ZD{4}\}.
	\end{equation}
	This codespace has logical Pauli operators
	\begin{equation}
	\ZL = \ZD{1}\ZD{2},\ \ZD{3}\ZD{4},\ \ZD{1}\ZD{4},\ \mathrm{and}\ \ZD{2}\ZD{3},
    \label{eq:logicalZ}
	\end{equation}
	\begin{equation}
	\XL = \XD{1}\XD{3}\ \mathrm{and}\ \XD{2}\XD{4},
    \label{eq:logicalX}
	\end{equation}
	that anti-commute with each other and commute with $\mathscr{S}$, and logical computational basis
	\begin{align}
	\zeroL &= \frac{1}{\sqrt2} \left( \ket{0000} + \ket{1111} \right),\\
	\oneL  &= \frac{1}{\sqrt2} \left( \ket{0101} + \ket{1010} \right).
	\end{align}
	Measuring the stabilizers using three ancilla qubits ($\Aone$, $\Atwo$ and $\Athree$ in Fig.~\ref{fig:state_tomo}a) allows detection of all individual physical-qubit errors. Such errors change the outcome of one or more stabilizers to $m=-1$. However, no error syndrome combination is unique to a single error. For instance, a phase flip in any one data qubit triggers the same syndrome: $m_\mathrm{A2}=-1$. Consequently, this code cannot be used to correct such errors.
    We thus perform state stabilization by post-selecting runs in which no error is detected by the stabilizer measurements in any cycle.
	In this error-detection context, an operation is fault-tolerant if any single-fault produces a non-trivial syndrome and can therefore be post-selected out~\cite{Tomita14}.
\section{Results} \label{sec:results}
	\subsection{Stabilizer measurements}
		Achieving high performance in a code hinges on performing projective quantum parity (stabilizer) measurements with high assignment fidelity and low additional  backaction.
		We implement each of the stabilizers in $\mathscr{S}$ using a standard indirect-measurement scheme~\cite{Saira14, Takita16} with a dedicated ancilla. As a fidelity metric, we measure the average probability to correctly assign the parity $\ZD{1}\ZD{3}$, $\ZD{1}\ZD{2}\ZD{3}\ZD{4}$ and $\ZD{1}\ZD{3}$ of physical computational states of the data-qubit register, finding $94.2\%$, $86.1\%$ and $97.2\%$, respectively (see Fig.~\ref{fig:parity_checks}).
	
	\subsection{Logical state initialization using stabilizer measurements}
		A practical means to quantify the backaction of stabilizer measurements is using them to initialize logical states.
		As proposed in Ref.~\onlinecite{Andersen20}, we can prepare arbitrary logical states by first initializing the data-qubit register in the product state
        \begin{equation}
		\ket{\psi}=\left(C_{\theta/2}\ket{0}+S_{\theta/2}\ket{1}\right)\ket{0}\left(C_{\theta/2}\ket{0}+S_{\theta/2}e^{i\phi}\ket{1}\right)\ket{0}
		\label{eq:state_prep}
		\end{equation}
        using single-qubit rotations $R_{y}^\theta$ on $\Done$ and $R_{\phi}^\theta$ on $\Dthree$ acting on $\ket{0000}$ ($C_{\alpha}=\cos\alpha$ and $S_{\alpha}=\sin\alpha$).
        A follow-up round of stabilizer measurements ideally projects the four-qubit state onto the logical state
		\begin{equation}
		\ket{\psi_\mathrm{L}}= \left(C_{\theta/2}^2\zeroL+S_{\theta/2}^2e^{i\phi}\oneL\right)\;/\;\sqrt{C_{\theta/2}^4+S_{\theta/2}^4}
		\label{eq:state_prep2}
		\end{equation}
		with probability
        \begin{equation}
        P=\frac{1}{2}\left(C_{\theta/2}^4+S_{\theta/2}^4\right).
        \label{eq:Prob}
        \end{equation}
		We use this procedure to target initialization of the logical cardinal states $\zeroL$, $\oneL$, $\plusL=\big(\zeroL+\oneL\big)/\sqrt{2}$, and $\minusL=\big(\zeroL-\oneL\big)/\sqrt{2}$.
		For the first two states, the procedure is fault-tolerant according to the definition above. We characterize the produced states using full four-qubit state tomography including readout calibration and maximum-likelihood estimation (MLE) (Fig.~\ref{fig:state_tomo}). The fidelity $\FidfourQ$ to the ideal four-qubit target states is $90.0\%$, $92.9\%$, $77.80\%$, and $77.09\%$, respectively.  For each state, we can extract a logical fidelity $\FidL$ by further projecting the obtained four-qubit density matrix onto the codespace~\cite{Andersen20}, finding $99.83\%$, $99.97\%$, $97.02\%$, and $95.54\%$, respectively (see Methods). This sharp increase from $\FidfourQ$ to $\FidL$ demonstrates that the vast majority of errors introduced by the parity check are weight-1 and detectable. A simple modification makes the initialization of $\plusL$ ($\minusL$) also fault-tolerant: initialize the data-qubit register in a different product state, namely  $\ket{+\!+\!++}$ ($\ket{+\!+\!--}$), before performing the stabilizer measurements. With this modification, $\FidfourQ$ increases to $85.4\%$ ($84.6\%$) and $\FidL$ to $99.78\%$ ($99.64\%$), matching the performance achieved when targetting $\zeroL$ and $\oneL$.

	\subsection{Logical measurement of arbitrary states}
		\begin{figure}
		    \centering
		    \includegraphics[width=0.5\textwidth]{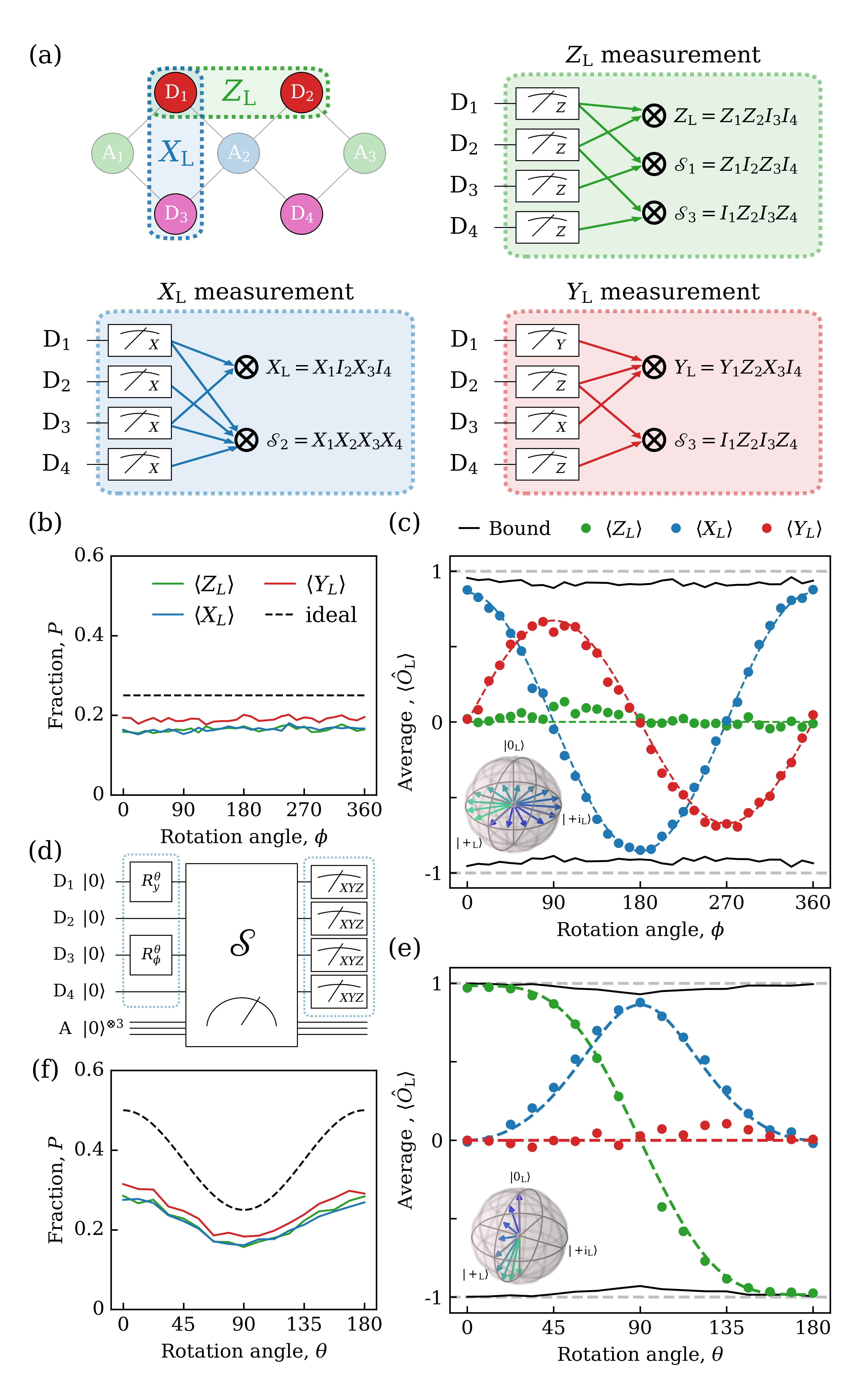}
		    \caption{\textbf{Arbitrary logical-state initialization and measurement in the logical cardinal bases.}
				    (a) Assembly of data-qubit measurements used to evaluate logical operators $\ZL$, $\XL$ and $\YL$ with additional error detection.
				    (d) Initialization of logical states using the procedure described in Eq.~\ref{eq:state_prep}.
				    (c, e) $\ZL$, $\XL$ and $\YL$ logical measurement results as a function of the gate angles $\phi$ (c) and $\theta$ (e). The colored dashed curves show a fit of the analytical prediction based on Eqs.~\ref{eq:stateZ} and~\ref{eq:stateY} to the data and the dark curve denotes a bound based on the measured $\FidL$ of each state.
				    (b, f) Total fraction $P$ of post-selected data as a function of the input angle for each logical measurement. The dashed curve shows the ideal fraction given by Eq.~\ref{eq:Prob}.
				    }
		    \label{fig:Arb_prep}
		\end{figure}
		A key feature of a code is the ability to measure logical operators.
        In the surface code, we can measure $\XL$ ($\ZL$) fault-tolerantly, albeit destructively, by simultaneously measuring all data qubits in the $X$ ($Z$) basis to obtain a string of data-qubit outcomes (each $+1$ or $-1$). The value assigned to the logical operator is the computed product of data-qubit outcomes as prescribed by Eq.~\ref{eq:logicalX} (\ref{eq:logicalZ}). Additionally, the outcome string is used to compute a value for the stabilizer(s) $\XD{1}\XD{2}\XD{3}\XD{4}$  ($\ZD{1}\ZD{3}$ and $\ZD{2}\ZD{4}$), enabling a final step of error detection (Fig.~\ref{fig:Arb_prep}a). Measurement of $\YL=+i\XL\ZL=\YD{1}\ZD{2}\XD{3}$ is not fault-tolerant. However, we lower the logical assignment error by also measuring  $\Dfour$ in the $Z$ basis to compute a value for $\ZD{2}\ZD{4}$ and thereby detect bit-flip errors in $\Dtwo$ and $\Dfour$.
		
        We demonstrate $\ZL$, $\XL$ and $\YL$ measurements on logical states prepared on two orthogonal planes of the logical Bloch sphere.
        Setting $\theta=\pi/2$ and sweeping $\phi$, we ideally prepare logical states on the equator (Fig.~\ref{fig:Arb_prep}d)
		\begin{equation}
		\ket{\psi_\mathrm{L}} = \big(\zeroL + e^{i\phi}\oneL\big)\;/\;\sqrt{2}.
		\label{eq:stateZ}
		\end{equation}
		We measure the produced states in the $\ZL$, $\XL$ and $\YL$ bases and obtain experimental averages $\avgZL$, $\avgXL$ and $\avgYL$. 
        As expected, we observe sinusoidal oscillations in $\avgXL$ and $\avgYL$ and near-zero $\avgZL$. 
        We extract logical assignment fidelities $\FidLReadout$ for $\XL$ and $\YL$ from the amplitude of the oscillations and separating the effect of initialization error: 
        \begin{equation}
        	(2\FidLReadout-1)(2\FidL-1)=\mathrm{max}|\langle O_\mathrm{L}\rangle|\:,\: O\in\{X, Y\}.
        \end{equation}
        We find $\FidLReadout=95.8\%$ for $\XL$ and $87.5\%$ for $\YL$, which manifests the non-fault-tolerant nature of $\YL$ measurement. A second manifestation is the higher fraction $P$ of post-selected data in this case (Fig.~\ref{fig:Arb_prep}b).
         
		Setting $\phi=0$ and sweeping $\theta$, we then prepare logical states on the $\XL$-$\ZL$ plane of the logical Bloch sphere (Fig.~\ref{fig:Arb_prep}d), ideally
		\begin{equation}
		\ket{\psi_\mathrm{L}} = \left(C_{\theta/2}^2\zeroL + S_{\theta/2}^2\oneL\right)\;/\;\sqrt{C_{\theta/2}^4+S_{\theta/2}^4}.
		\label{eq:stateY}
		\end{equation}
        Note that due to the changing overlap of the initial product state with the codespace, $P$ is now a function of $\theta$ (Eq.~\ref{eq:Prob}). 
        Using the same procedure as above, we extract $\FidLReadout=99.4\%$ for $\ZL$ and $96.4\%$ for $\XL$. 
        Although both measurements are fault-tolerant, $\FidLReadout$ is higher for $\ZL$. This arises because the $\ZL$ measurement is only vulnerable to vertical double bit-flip errors while the $\XL$ measurement is vulnerable to horizontal and diagonal double phase-flip errors. 

	\subsection{Logical gates}
		\begin{figure}
		\centering
		    \includegraphics[width=0.4355\textwidth]{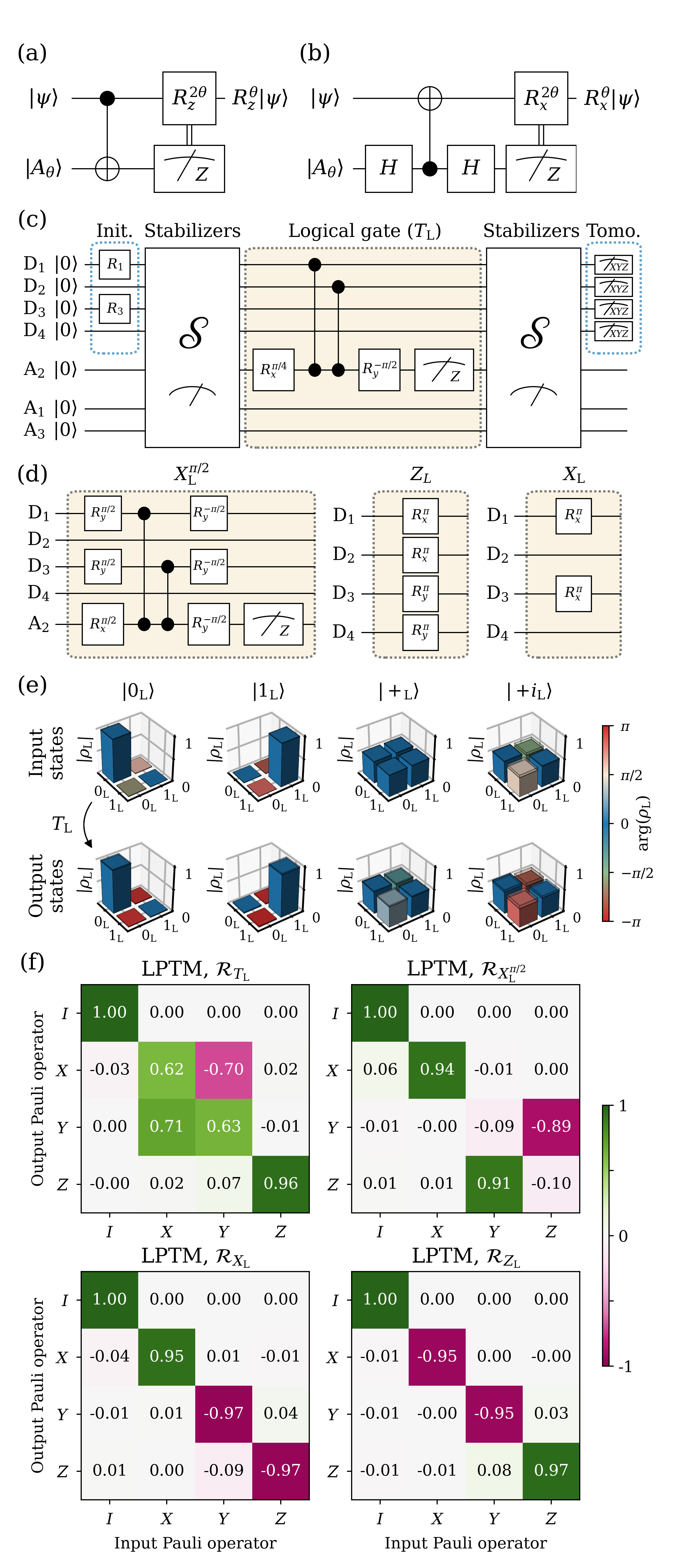}
		    \caption{\textbf{Logical gates and their characterization.} (a, b) General gate-by-measurement schemes realizing arbitrary rotations around the $Z$ (a) and $X$ (b) axis of the Bloch sphere. (c) Process tomography experiment of the $\TL$ gate. Input cardinal logical  states are initialized using the method of Fig.~\ref{fig:Arb_prep}. Output states are measured following a second round of stabilizer measurements. (d) Logical $\XL^{\pi/2}$, $\ZL$ and $\XL$ gates compiled using the hardware-native gateset. (e) Logical state tomography of input and output states of the $\TL$ gate. These logical density matrices are obtained by performing four-qubit tomography of the data qubits and then projecting onto the codespace. (f) Extracted logical Pauli transfer matrices.}
		    \label{fig:Logical_ops}
		\end{figure}
		\begin{figure*}[!htb]
		    \centering
		    \includegraphics[width=\textwidth]{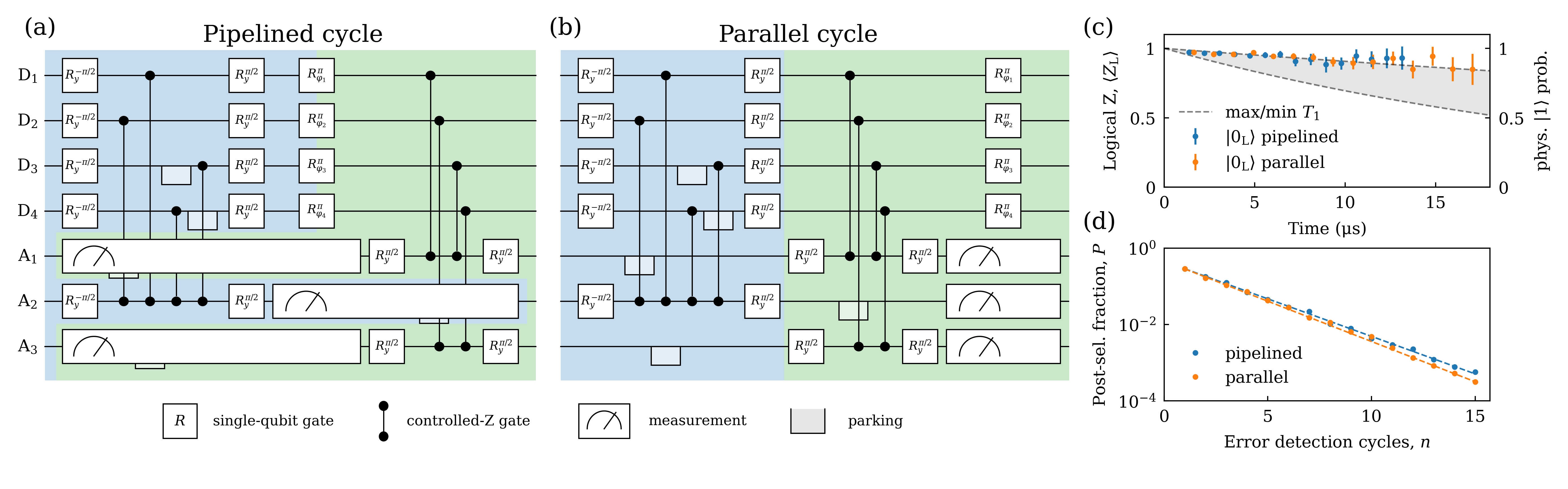}
		    \caption{\textbf{Repetitive error detection using pipelined and parallel stabilizer measurement schemes.} (a, b) Gate sequences used to implement the pipelined (a) and parallel (b) stabilizer measurement schemes. Gate duration is $20~\ns$ for single-qubit gates, $60~\ns$ for controlled-Z (CZ) gates and parking~\cite{Versluis17, Andersen19}, and $540~\ns$ for ancilla readout. The order of CZs in the $\XD{1}\XD{2}\XD{3}\XD{4}$ stabilizer (blue shaded region) prevents the propagation of ancilla errors into logical qubit errors~\cite{Tomita14}. The total cycle duration for the pipelined (parallel) scheme is $840~\ns$ ($1000~\ns$). (c) Estimated $\ZL$ expectation value, $\langle \ZL\rangle$, measured for the $\zeroL$ state versus the duration of the experiment using the pipelined (blue) and the parallel (orange) schemes. We also plot the excited-state probability (right axis) set by the maximum and minimum physical qubit $\Tone$. 
		    (d) Post-selected fraction of data versus the number of error detection cycles $n$ for the pipelined (blue) and parallel (orange) scheme.}
		    \label{fig:state_stab}
		\end{figure*}
		Finally, we demonstrate a suite of gates enabling universal logical-qubit control (Fig.~\ref{fig:Logical_ops}).
		Full control of the logical qubit requires a gate set comprising Clifford and non-Clifford logical gates.
		Some Clifford gates, like $\ZL$ and $\XL$, can be implemented transversally and therefore fault-tolerantly (Fig.~\ref{fig:Logical_ops}d). 
        We perform arbitrary rotations (generally non-fault-tolerant) about the $\ZL$ axis using the standard gate-by-measurement circuit~\cite{Aliferis05} shown in Fig.~\ref{fig:Logical_ops}a. In our case, the ancilla is physical ($\Atwo$), while the qubit transformed is our logical qubit. 
        The rotation angle $\theta$ is set by the initial ancilla state $\ket{A_\theta}=(\ket{0}+e^{i\theta}\ket{1})/\sqrt{2}$. 
        Since we cannot do binary-controlled $\ZL$ rotations, we simply post-select runs in which the measurement outcome is $m_\mathrm{A2}=+1$. 
        Choosing $\theta=\pi/4$ implements the non-Clifford $\TL=\ZL^{\pi/4}$ gate. A similar circuit (Fig.~\ref{fig:Logical_ops}b) can be used to perform arbitrary rotations around the $\XL$ axis. 
        We compile both circuits using our hardware-native gateset (Figs.~\ref{fig:Logical_ops}c,d).
		To assess logical-gate performance, we perform logical process tomography using the procedure illustrated in Fig.~\ref{fig:Logical_ops}e for $\TL$. First, we initialize into each of the six logical cardinal states $\{\zeroL,\oneL,\plusL,\minusL,\iplusL,\iminusL\}$.
		We characterize each actual input state by four-qubit state tomography and project to the codespace to obtain a logical density matrix. Next, we similarly characterize each output state produced by the logical gate and a second round of stabilizer measurements to detect errors occurred in the gate.
		Using this over-complete set of input-output logical-state pairs, combined with MLE (see Methods), we extract a logical Pauli transfer matrix (LPTM). 
        The resulting LPTMs for  the non-fault-tolerant $\TL$ and $\XL^{\pi/2}$ gates as well as the fault-tolerant $\ZL$ and $\XL$ are shown in Fig.~\ref{fig:Logical_ops}e. From the LPTMs, we extract average logical gate fidelities $\FidLGate$ (Eq.~\ref{eq:Favg}) 97.3\%, 95.6\%, 97.9\%, and 98.1\%, respectively.

	\subsection{Pipelined versus parallel stabilizer measurements}
		A scalable control scheme is fundamental to realize surface codes with large code distance.
		To this end, we now compare the performance of two schemes suitable for the quantum hardware architecture proposed in Ref.~\onlinecite{Versluis17}.
		These schemes are scalable in the sense that their cycle duration remains independent of code distance.
		The pipelined scheme interleaves the coherent operations and ancilla readout steps associated with stabilizer measurements of type $X$ and $Z$ by performing the coherent operations of $X$ ($Z$) type stabilizers during the readout of $Z$ ($X$) type stabilizers. (Fig.~\ref{fig:state_stab}a).
		The parallel scheme performs all ancilla readouts simultaneously (Fig.~\ref{fig:state_stab}b).
		To compare their performance, we initialize and stabilize $\zeroL$ for up to $n=15$ cycles.
		We separately calibrate the equatorial rotation axis of refocusing pulses ($R_{\varphi_i}^\pi$) in each scheme to extract the best performance in both schemes.
		At each $n$, we take data back-to-back for the two schemes in order to minimize the effect of parameter drift, repeating each experiment up to $256\times10^3$ times.
		Figure~\ref{fig:state_stab}c shows the $\ZL$ measurement outcome averaged over the post-selected runs.
		We extract the error-detection rate $\gamma$ from the $n$-dependence of the fraction of post-selected data $P$ (Fig.~\ref{fig:state_stab}d) using the procedure described in Methods.
		We observe that the error rate is slightly lower for the pipelined scheme ($\Gpip\sim 45\%$), most likely due to the shorter duration of the cycle. This superiority is consistent across different input logical states (see Fig.~\ref{fig:Pip_vs_par}) with an average ratio $\Gpip/\Gpar \sim 97\%$. 
\section{Discussion} \label{sec:discussion}

We have demonstrated a suite of logical-level initialization, gate and measurement operations in a distance-2 superconducting surface code undergoing repetitive stabilizer measurements.
For each type of logical operation, we have quantified the increased performance of fault-tolerant variants over non-fault-tolerant variants. Table~\ref{tab:Log_Ops} summarizes all the results.
We can initialize the logical qubit to any point on the logical Bloch sphere, with logical fidelity surpassing Ref.~\onlinecite{Andersen20}.
In addition to characterizing initialized states using full four-qubit tomography, we also demonstrate logical measurements in all logical cardinal bases.
Finally, we demonstrate a universal single-qubit set of logical gates by performing logical process tomography, introducing the concept of a logical-level Pauli transfer matrix.

With a view towards implementing higher-distance surface codes using our quantum-hardware architecture~\cite{Versluis17}, we have compared the performance of two scalable stabilization schemes: the pipelined and parallel measurement schemes.
In this comparison, two main factors compete. On one hand, the shorter cycle time favors pipelining. On the other, the pipelining introduces extra dephasing on ancilla qubits of one type during readout of the other.
The performance of both schemes is comparable, but slightly higher for the pipelined scheme. From density-matrix simulations discussed in detail in the Supplementary Material,  we further understand that conventional qubit errors such as energy relaxation, dephasing and readout assignment error alone do not account for the net error-detection rate observed in the experiment (Fig.~\ref{fig:postsel_frac_model_comparison}). We believe that the dominant error source is instead leakage to higher transmon states incurred during CZ gates.
Our data (Fig.~\ref{fig:q_leak_histograms}) shows that the error detection scheme successfully post-selects leakage errors in both the ancilla and data qubits. Learning to identify these non-qubit errors and to correct them without post-selection is the subject of ongoing research~\cite{Varbanov20,Mcewen21,Battistel21} and an outstanding challenge in the quest for quantum fault-tolerance with higher-distance superconducting surface codes.

\begin{table}
    \resizebox{.48\textwidth}{!}{
        \begin{tabular}{c|c|c|c|c}
\hline
\multicolumn{2}{l|}{Logical operation}                      & Characteristic  & Logical fidelity metric           & value (\%)  \\
\hline \hline
\multirow{4}{*}{\rotatebox{90}{Init.       }} & $\zeroL$    & FT              & \multirow{4}{*}{$\FidL$}          & 99.83       \\
                                              & $\oneL$     & FT              &                                   & 99.97       \\
                                              & $\plusL$    & Non-FT/FT       &                                   & 97.02/99.78 \\
                                              & $\minusL$   & Non-FT/FT       &                                   & 95.54/99.64 \\
\hline
\multirow{3}{*}{\rotatebox{90}{Meas.       }} & $\ZL$       & FT              & \multirow{3}{*}{$\FidLReadout$}   & 99.4        \\
                                              & $\XL$       & FT              &                                   & $96.0^*$    \\
                                              & $\YL$       & Non-FT          &                                   & 87.5        \\
\hline
\multirow{4}{*}{\rotatebox{90}{Gate$\:\: $}}  & $\ZL$        & FT             & \multirow{4}{*}{$\FidLGate$}      & 98.1        \\
                                              & $\XL$        & FT             &                                   & 97.9        \\
                                              & $\XL^{\pi/2}$& Non-FT         &                                   & 95.6        \\
                                              & $\TL$        & Non-FT         &                                   & 97.3        \\
\hline
        \end{tabular}
    }
    \caption{\textbf{Summary of logical initialization, measurement, and gate operations and their performance.} 
    Fault-tolerant operations are labelled FT and non-fault tolerant ones Non-FT. $^*$Weighted average of values extracted from Figs.~\ref{fig:Arb_prep}c,d. 
    }
    \label{tab:Log_Ops}
\end{table}

\section{Methods} \label{sec:methods}
	\subsection{Device}
		We use a seven-transmon superconduting processor (Fig.~\ref{fig:state_tomo}b) featuring the quantum-hardware architecture proposed in Ref.~\onlinecite{Versluis17}. We employ flux-tunable transmons arranged in three frequency groups: a high-frequency group for $\Done$ and $\Dtwo$; a middle-frequency group for $\Aone$, $\Atwo$ and $\Athree$; and a low-frequency group for $\Dthree$ and $\Dfour$. Each transmon is transversely coupled to its nearest neighbor using a dedicated coupling bus resonator and features an individual microwave drive line for single-qubit gates, a flux line for two-qubit gates,  and a dispersively coupled readout resonator with Purcell filter for readout~\cite{Heinsoo18, Bultink20}. All transmons are flux biased to their maximal frequency (i.e., flux sweetspot~\cite{Schreier08}). Qubit relaxation ($\Tone$) and dephasing ($\Ttwoecho$) times lie in the range \mbox{27---102 $\us$} and \mbox{55---117 $\us$}, respectively. Detailed information on the implementation and performance of single- and two-qubit gates can be found in Ref.~\onlinecite{Negirneac20}. Device characteristics are also summarized in Table~\ref{tab:Device}.

	\subsection{State tomography}
		To perform state tomography on the prepared logical states, we measure the $4^4-1$ expectation values of data-qubit Pauli observables, \mbox{$p_i=\langle \sigma_i\rangle, \sigma_i \in \{I, X, Y, Z\}^{\otimes4}$} (except~$I^{\otimes 4}$). These are used to construct the density matrix
		\begin{equation}
			\rho = \sum_{i=0}^{4^4-1} \frac{p_i\sigma_i}{2^4},
		\end{equation}
		with $p_0=1$ corresponding to $\sigma_0=I^{\otimes 4}$.
		Due to statistical uncertainty in the measurement, the constructed state, $\rho$, might lack the physicality characteristic of a density matrix, that is, $\mathrm{Tr}(\rho)=1$ and $\rho\ge 0$.
		Specifically, $\rho$ might not satisfy the latter constraint, while the former is automatically satisfied by $p_0=1$.
		To enforce these constraints, we use a maximum-likelihood method~\cite{Chow12} to find the physical density matrix, $\rho_{\mathrm{ph}}$, that is closest to the measured state, where closeness is defined in terms of best matching the measurement results.
		We thus minimize the cost function $\sum_{i=0}^{4^4-1} |p_i - \mathrm{Tr}(\rho_{\mathrm{ph}}\sigma_i)|^2$, subject to $\mathrm{Tr}(\rho_{\mathrm{ph}})=1$ and $\rho_{\mathrm{ph}}\ge 0$.
		We find the optimal $\rho_{\mathrm{ph}}^{\mathrm{opt}}$ using the convex-optimization package \emph{cvxpy} via \emph{cvx-fit} in Qiskit~\cite{Qiskit19}.
		The fidelity to a target pure state, $\ket{\psi}$, is then computed as
		\begin{equation}
			F= \bra{\psi}\rho_{\mathrm{ph}}^{\mathrm{opt}}\ket{\psi}.
			\label{eq:state_fidelity}
		\end{equation}
		One can further project $\rho_{\mathrm{ph}}$ onto the codespace to obtain a logical state $\rho_\mathrm{L}$ using
		\begin{equation}
			\rho_\mathrm{L} = \frac{1}{2}\sum_{i} \frac{\mathrm{Tr}(\rho_\mathrm{ph}\sigma^\mathrm{L}_i)}{\mathrm{Tr}(\rho_\mathrm{ph} I_\mathrm{L})}\sigma_i^\mathrm{L}\:\:,\:\: \sigma^\mathrm{L}_i \in \{I_\mathrm{L},X_\mathrm{L}, Y_\mathrm{L}, Z_\mathrm{L}\},
		\end{equation}
		where $I_\mathrm{L}$ is the projector onto the codespace. Here, we can compute the logical fidelity $\FidL$ using Eq.~\ref{eq:state_fidelity}.

	\subsection{Process tomography in the codespace}
		A general single-qubit gate can be described~\cite{Chow12}  by a Pauli transfer matrix (PTM) $\R$ that maps an input state described by \mbox{$p_i=\langle\sigma_i\rangle,\sigma_i\in\{I, X, Y, Z\}$}, with $p_0=1$, to an output state $p'$:
		\begin{equation}
			p'_j=\sum_{i}\R_{ij}p_i.
		\end{equation}
		To construct $\R$ in the codespace, we use an overcomplete set of input states, \mbox{$\{\zeroL,\oneL,\plusL,\minusL,\iplusL,\iminusL\}$}, and their corresponding output states and perform linear inversion. The input and output logical states are characterized using state tomography of the data qubits to find the four-qubit state $\rho$, which is then projected to the codespace using:
		\begin{equation}
			p^\mathrm{L}_i = \frac{\mathrm{Tr}(\rho\sigma^\mathrm{L}_i)}{\mathrm{Tr}(\rho I_\mathrm{L})}\:\:,\:\: \sigma^\mathrm{L}_i \in \{I_\mathrm{L},X_\mathrm{L}, Y_\mathrm{L}, Z_\mathrm{L}\},
		\end{equation}
		We find that all the measured logical states already satisfy the constraints of a physical density matrix.
		This is likely to happen as one-qubit states that are not very pure usually lie within the Bloch sphere even within the uncertainty in the measurement.
		The constructed LPTM, however, might not satisfy the constraints of a physical quantum channel, that is, trace preservation and complete positivity (TPCP).
		These are better expressed by switching from the PTM representation to the Choi representation. The Choi state $\rho^{\R}$ can be computed as
		\begin{equation}
			\rho^{\R} = \frac{1}{4} \sum_{i,j} \R_{ij} \, \sigma_j^T \otimes \sigma_i,
		\end{equation}
		where the first tensor-product factor corresponds to an auxiliary subsystem.
		The TPCP constraints are $\mathrm{Tr}(\rho^{\R}_\mathrm{ph})=1$, $\rho^{\R}_\mathrm{ph}\ge 0$ and $\mathrm{Tr}_1(\rho^{\R}_\mathrm{ph})=1/2$, where $\mathrm{Tr}_1$ is the partial trace over the auxiliary subsystem.
		In other words, $\rho^{\R}_\mathrm{ph}$ is a density matrix satisfying an extra constraint.
		We then find the optimal $\rho^{\R,\mathrm{opt}}_\mathrm{ph}$ using the same convex-optimization methods as for state tomography and adding this extra constraint~\cite{Chow12,DeJong19}.
		We compute the corresponding LPTM via
		\begin{equation}
			(\R^{\mathrm{opt}}_\mathrm{ph})_{ij} = \mathrm{Tr}(\rho^{\R,\mathrm{opt}}_\mathrm{ph} \, \sigma_j^T\otimes \sigma_i)
		\end{equation}
		and the average logical gate fidelity using
		\begin{equation}
			\FidLGate=\frac{\mathrm{Tr}(\R_\mathrm{ideal}^\dagger \R^{\mathrm{opt}}_\mathrm{ph})+2}{6},
			\label{eq:Favg}
		\end{equation}
		where $\R_\mathrm{ideal}$ is the LPTM of the ideal target gate.

	\subsection{Extraction of error-detection rate}
		The fraction of post-selected data $P$ in the repetitive error detection experiment (Fig.~\ref{fig:state_stab}b) decays exponentially with the number of cycles $n$. 
        This is consistent with a constant error-detection rate per cycle $\gamma$. We extract this rate by fitting the function
		\begin{equation}
			P(n) = A(1-\gamma)^n.
		\end{equation}

%

\end{bibunit}
\section*{Acknowledgements} \label{sec:acknowledgements}
    We thank R.~Sagastizabal, M.~Sarsby and T.~Stavenga for experimental assistance, and G.~Calusine and W.~Oliver for providing the traveling-wave parametric amplifiers used in the readout amplification chain. This research is supported by the Office of the Director of National Intelligence (ODNI), Intelligence Advanced Research Projects Activity (IARPA), via the U.S. Army Research Office Grant No. W911NF-16-1-0071, and by Intel Corporation. The views and conclusions contained herein are those of the authors and should not be interpreted as necessarily representing the official policies or endorsements, either expressed or implied, of the ODNI, IARPA, or the U.S. Government. B.~M.~V., F.~B. and B.~M.~T. are supported by ERC Grant EQEC No. 682726.

\section*{Author contributions}
	J.~F.~M. performed the experiment and data analysis.
	M.~B., N.~H. and L.~D.~C. designed the device.
	N.~M., C.~Z. and A.~B. fabricated the device.
	J.~F.~M. and H.~A. calibrated the device.
	M.~S.~M. and W.~V. designed the control electronics.
	B.~M.~V. performed the numerical simulations and F.~B. implemented the MLE method. 
    B.~M.~T. supervised the theory work.
	J.~F.~M. and L.~D.~C. wrote the manuscript with contributions from B.~M.~V., F.~B. and B.~M.~T., and feedback from all coauthors.
	L.~D.~C. supervised the project. 
\section{Competing Interests}
	The authors declare no competing interests.
\onecolumngrid
\clearpage
\renewcommand{\theequation}{S\arabic{equation}}
\renewcommand{\thefigure}{S\arabic{figure}}
\renewcommand{\thetable}{S\arabic{table}}
\renewcommand{\bibnumfmt}[1]{[S#1]}
\renewcommand{\citenumfont}[1]{S#1}
\setcounter{figure}{0}
\setcounter{equation}{0}
\setcounter{table}{0}
\begin{bibunit}[apsrev4-2]
\section*{Supplemental material for 'Logical-qubit operations in an error-detecting surface code'}

This supplement provides additional information in support of statements and claims made in the main text.

\section{Device characteristics}
	\label{sec:one}
    \begin{table}[!h]
    \begin{tabular}{cccccccc}
    \hline
    Qubit               & $\Done$ & $\Dtwo$ & $\Dthree$ & $\Dfour$ &  $\Aone$  & $\Atwo$ & $\Athree$ \\
    \hline
    Qubit transition frequency at sweetspot, $\omega_q/2\pi$ (GHz) & 6.433 & 6.253 & 4.535 & 4.561 & 5.770 & 5.881 & 5.785 \\
    Transmon anharmonicity, $\alpha/2\pi$ (MHz)                    & -280  & ---   & -320  & ---   & -290  & -285  & ---   \\
    Readout frequency, $\omega_r/2\pi$ (GHz)                       & 7.493 & 7.384 & 6.913 & 6.645 & 7.226 & 7.058 & 7.101 \\
    Relaxation time, $\Tone$ ($\us$)                               & 27    & 44    & 32    & 102   & 38    & 58    & 43    \\
    Ramsey dephasing time, $\Ttwostar$ ($\us$)                     & 44    & 55    & 51    & 103   & 55    & 60    & 52    \\
    Echo dephasing time, $\Ttwoecho$ ($\us$)                       & 59    & 70    & 55    & 117   & 69    & 79    & 73    \\
    Best multiplexed readout fidelity, $F_{\mathrm{RO}}$, (\%)     & 98.6  & 98.9  & 96.0  & 96.5  & 98.6  & 94.2  & 98.9  \\
    \hline
    \end{tabular}
    \caption{Summary of frequency, coherence and readout parameters of the seven transmons. Coherence times are obtained using standard time-domain measurements~\cite{Krantz19}. Note that temporal fluctuations of several $\us$ are typical for these values. The multiplexed readout fidelity, $F_{\mathrm{RO}}$, is the average assignment fidelity~\cite{Bultink18} extracted from single-shot readout histograms after mitigating residual excitation using initialization by measurement and post-selection~\cite{Riste12,Walter17}.}
    \label{tab:Device}
    \end{table}

    \begin{figure}[!h]
	    \centering
	    \includegraphics[width=0.6\textwidth]{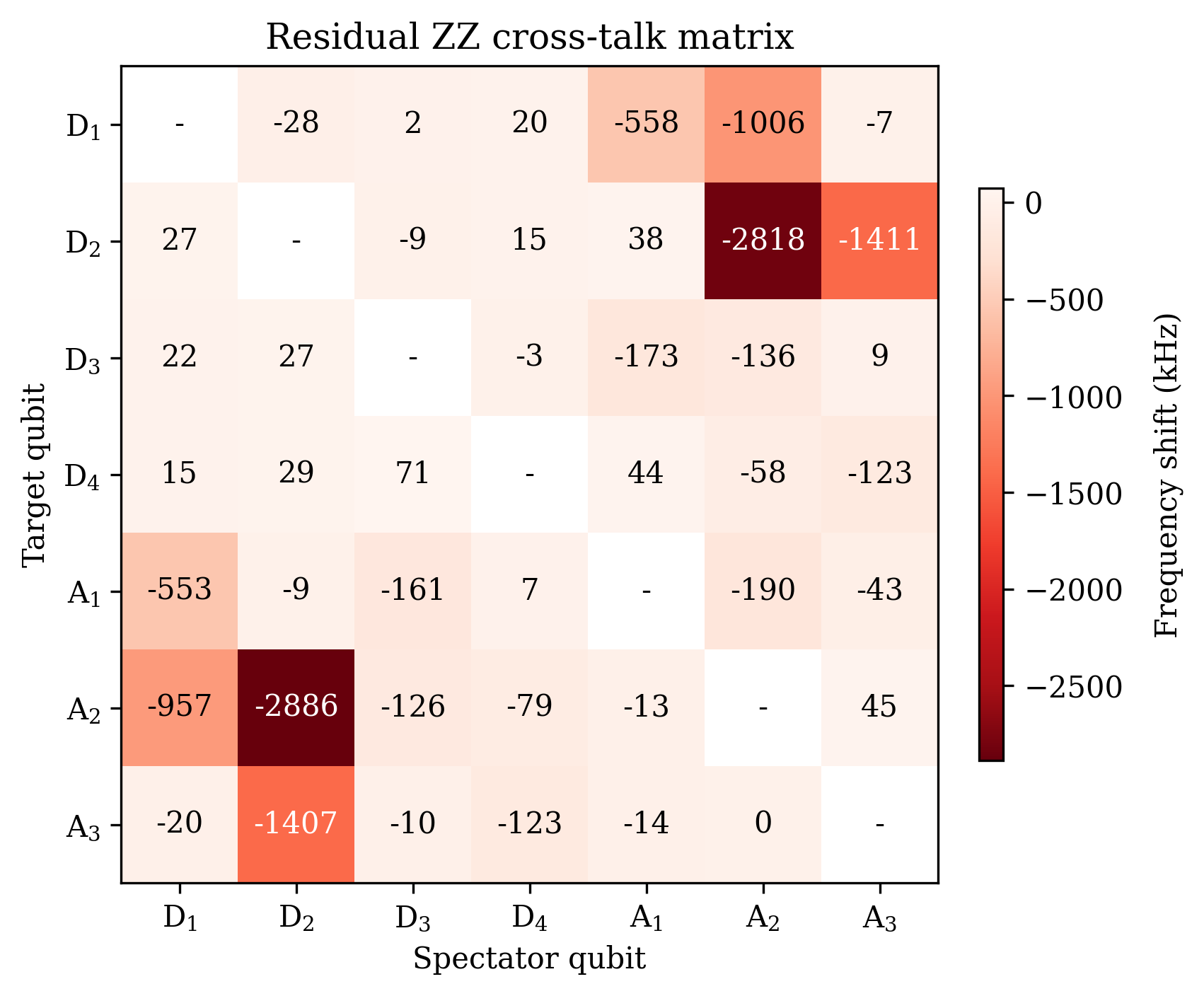}
	    \caption{Residual $ZZ$-coupling matrix.
        Measured residual $ZZ$ coupling between all transmon pairs at the bias point (their simultaneous flux sweetspot~\cite{Schreier08}). Each matrix element denotes the frequency shift that the target qubit experiences due to the spectator qubit being in the excited state, $\ket{1}$. The procedure used for this measurement is similar to the one described in Ref.~\onlinecite{Sagastizabal20}.}
	    \label{fig:zz_matrix}
	\end{figure}

\clearpage
\section{Parity-check performance}
	\begin{figure}[!htb]
	    \centering
	    \includegraphics[width=0.75\textwidth]{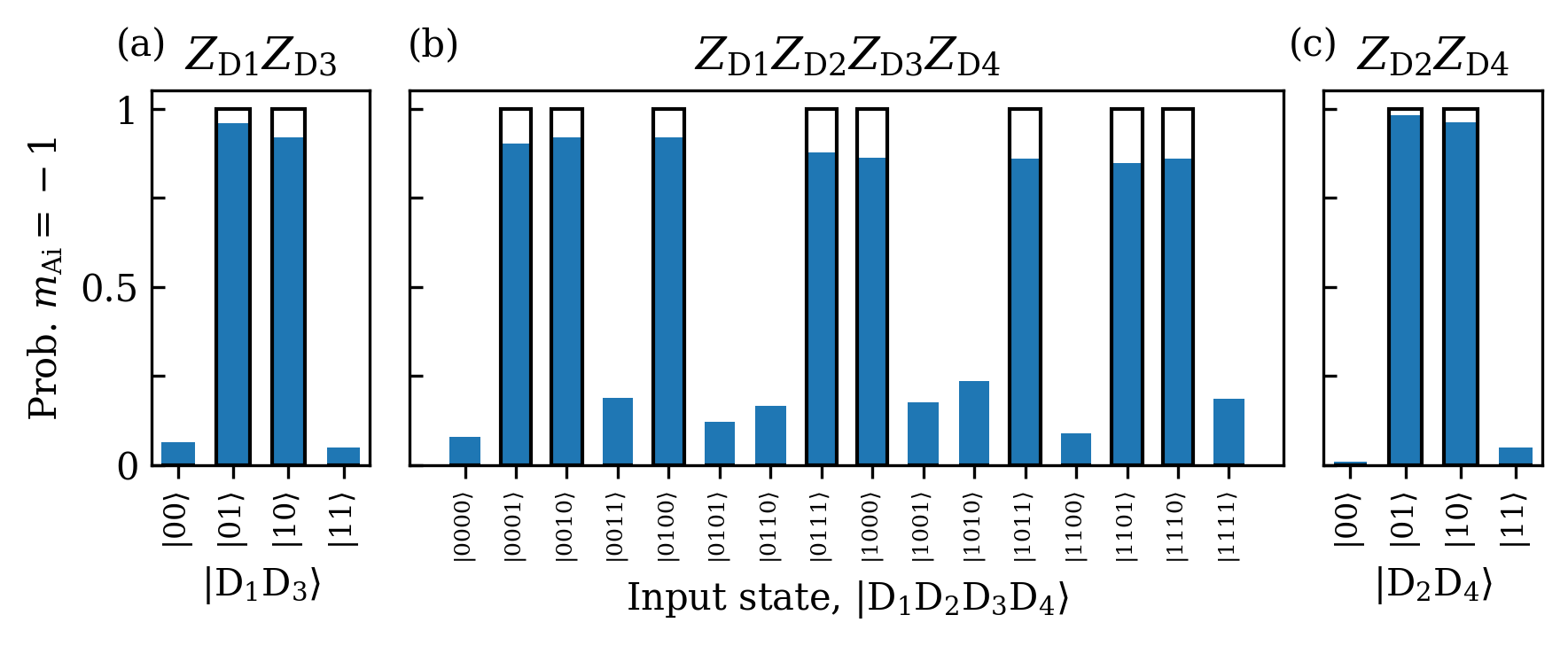}
	    \caption{Characterization of the assignment fidelity of $Z$-type parity checks (a) $\ZD{1}\ZD{3}$, (b) $\ZD{1}\ZD{2}\ZD{3}\ZD{4}$, and (c) $\ZD{2}\ZD{4}$  implemented using $\Zone$, $\X$, and $\Ztwo$, respectively.
        Each parity check is benchmarked by preparing the relevant data qubits in a computational state and then measuring the probability of ancilla outcome $m_{Ai}=-1$. 
        Measured (ideal) probabilities are shown as solid blue bars (black wireframe). From the measured probabilities we extract average assignment fidelities $94.2\%$, $86.1\%$ and $97.2\%$, respectively.}
	    \label{fig:parity_checks}
	\end{figure}

\section{State stabilization}

    \begin{figure}[!htb]
        \centering
        \includegraphics[width=0.95\textwidth]{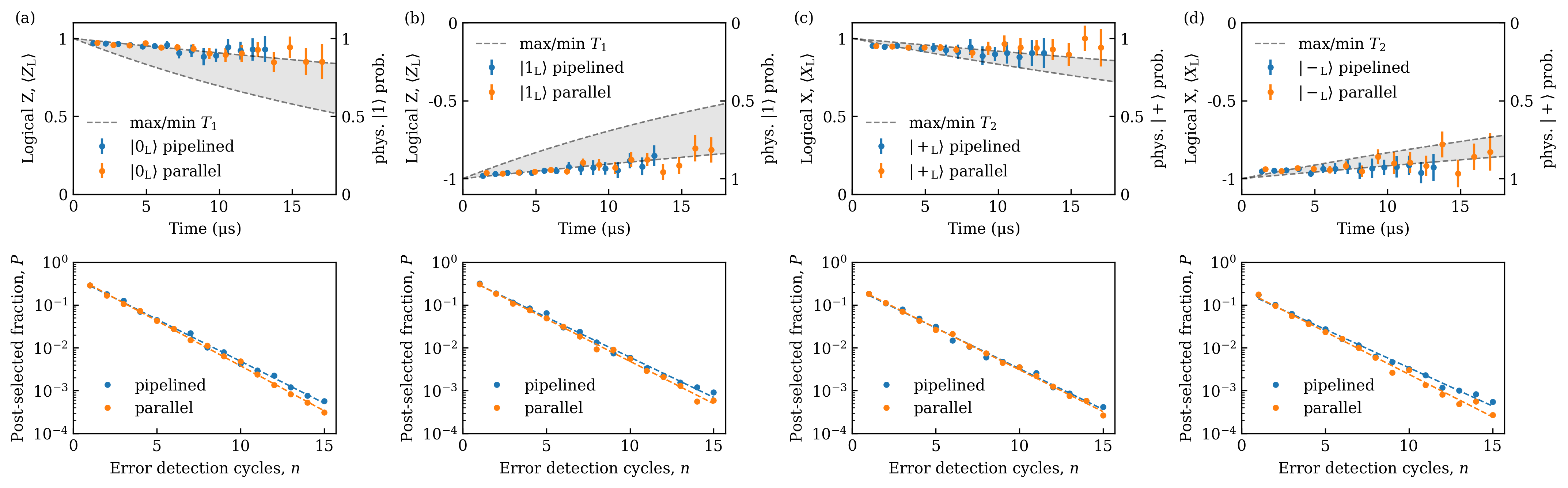}
        \caption{Stabilization of logical cardinal states by repetitive error detection using the pipelined and parallel schemes. From left to right, the stabilized logical states are $\zeroL$, $\oneL$, $\plusL$ and $\minusL$. For each logical state, the top panel shows the evolution of the relevant logical operator as a function of number of cycles, $n$, plotted versus wall-clock time. Error bars are estimated based on the statistical uncertainty given by $P(n)$. The shaded area indicates the range of physical qubit $\Tone$ values (a and b) and $\Ttwoecho$ values (c and d) plotted on the right-axis.
        Each bottom panel shows the corresponding post-selected fraction of data, $P(n)$. }
        \label{fig:Pip_vs_par}
    \end{figure}

\clearpage
\section{Numerical analysis}
\subsection{Leakage in experiment}
\begin{figure}[!h]
    \centering
    \includegraphics[width=0.8\textwidth]{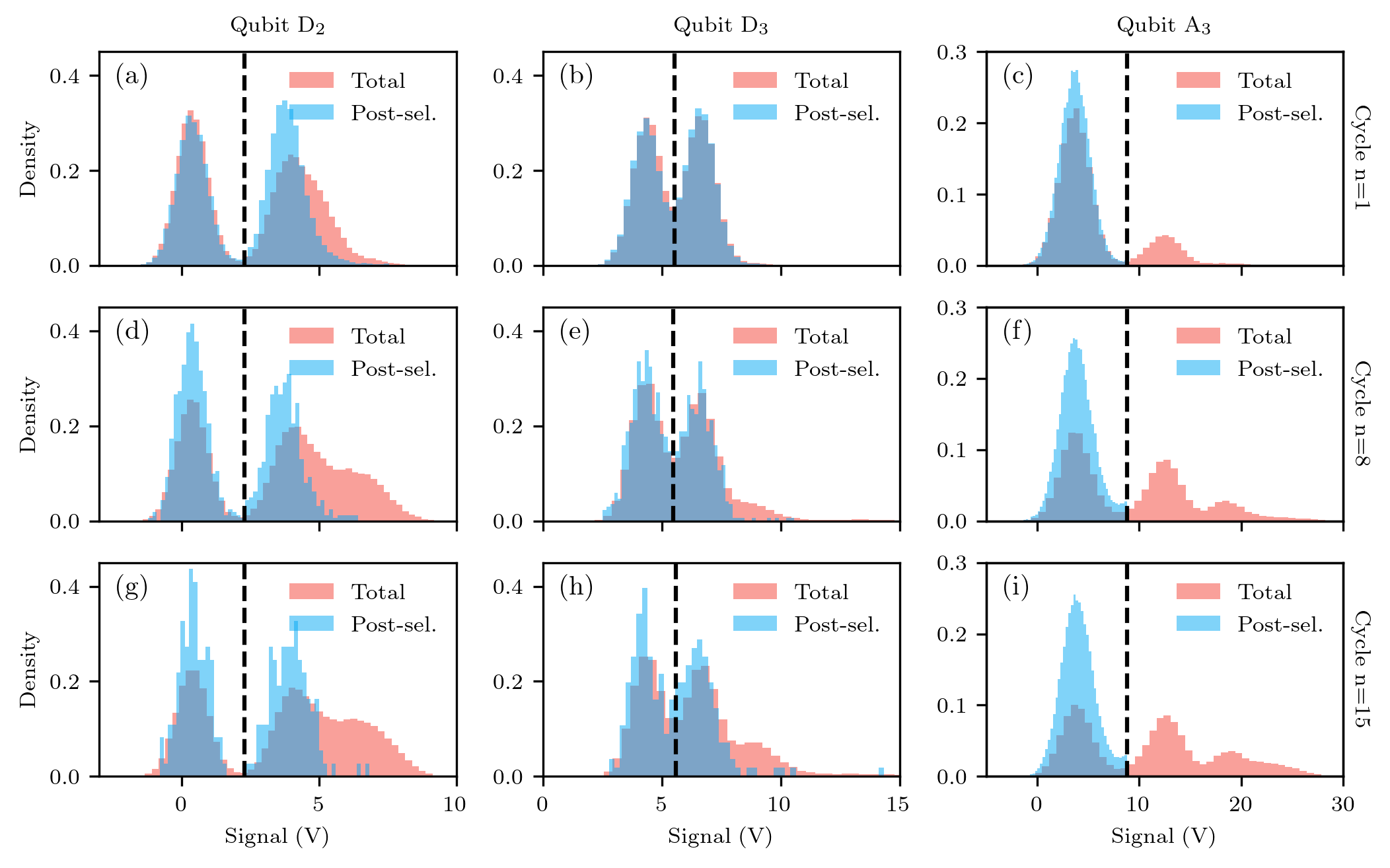}
    \caption{Single-shot readout histograms obtained at cycle $n$ over all shots (red) and the post-selected shots based on detecting no error in any cycles up to \(n\) (blue) for \(\Dtwo\) (left), \(\Dthree\) (middle) and \(\Ztwo\) (right) and at cycle $n=1$ (top row), $n=8$ (middle row) and $n=15$ (bottom row).
    The dashed black lines indicate the thresholds used to discriminate $\ket{0}$ from $\ket{1}$.}
    \label{fig:q_leak_histograms}
\end{figure}

We observe a clear signature of leakage accumulation with the increasing number of error-detection cycles in the single-shot readout histograms obtained at the end of each experiment.
In~\cref{fig:q_leak_histograms} we show examples of this accumulation for $\Dtwo$, $\Dthree$ and $\Ztwo$ at cycles $n=1$, $n=8$ and $n=15$.
For dispersive readout, a transmon in state~\(\ket{2}\) induces a different frequency shift in the readout resonator compared to state~\(\ket{0}\) or \(\ket{1}\).
The increased number of data points at $n=8$ and $n=15$ shown in~\cref{fig:q_leak_histograms}, following a Gaussian distribution with a mean and standard deviation different from those observed at $n=1$ is thus a clear manifestation of leakage to the higher-excited states (mostly to $\ket{2}$).
We believe that the dominant source of leakage in our processor are the CZ gates~\cite{Rol19,Negirneac20}.
However, the leakage rate $\leakrate$ for each gate has not been experimentally characterized, e.g., by performing leakage-modified randomized benchmarking experiments~\cite{Wood18,Asaad16}. This is because our CZ tune-up procedure is performed in a parity-check block unit. This maximizes the performance of the parity-check but makes the gate unfit for randomized benchmarking protocols.
We can estimate the population $p^{\mathcal{L}}\left(n\right)$ in the leakage subspace $\mathcal{L}$ at cycle $n$ from the single-shot readout histograms.
We perform a fit of a triple Gaussian model to the histograms from which we extract the voltage that allows for the best discrimination of $\ket{2}$ from $\ket{1}$ and $\ket{0}$.
The leaked population $p^{\mathcal{L}}\left(n\right)$ is then given by the fraction of shots declared as $\ket{2}$ over the total number of shots.
Assuming that leakage is only induced by the CZ gates (on the transmon being fluxed to perform the gate) and that each CZ gate has the same leakage rate $\leakrate$, we can use the Markovian model presented in Ref.~\onlinecite{Varbanov20} to estimate the $\leakrate$ value leading to the observed population \(p^{\mathcal{L}}\left(n\right)\).
This analysis gives a $\leakrate$ estimate in the approximate range $1-4\%$ for most transmons.
However, we do not consider these estimates to be accurate due to the low fidelity with which $\ket{2}$ can be distinguished from $\ket{1}$ and instead treat $\leakrate$ as a free parameter in our simulations (see below).

The histograms of the post-selected shots in~\cref{fig:q_leak_histograms} demonstrate that post-selection rejects runs where leakage on those transmons occurred.
Thus, while leakage may considerably impact the error-detection rate in the experiment~\cite{Varbanov20}, we do not expect it to significantly affect the fidelity of the logical initialization, and gates.

\subsection{Density-matrix simulations}
\begin{figure}[!h]
    \centering
    \includegraphics[width=0.75\textwidth]{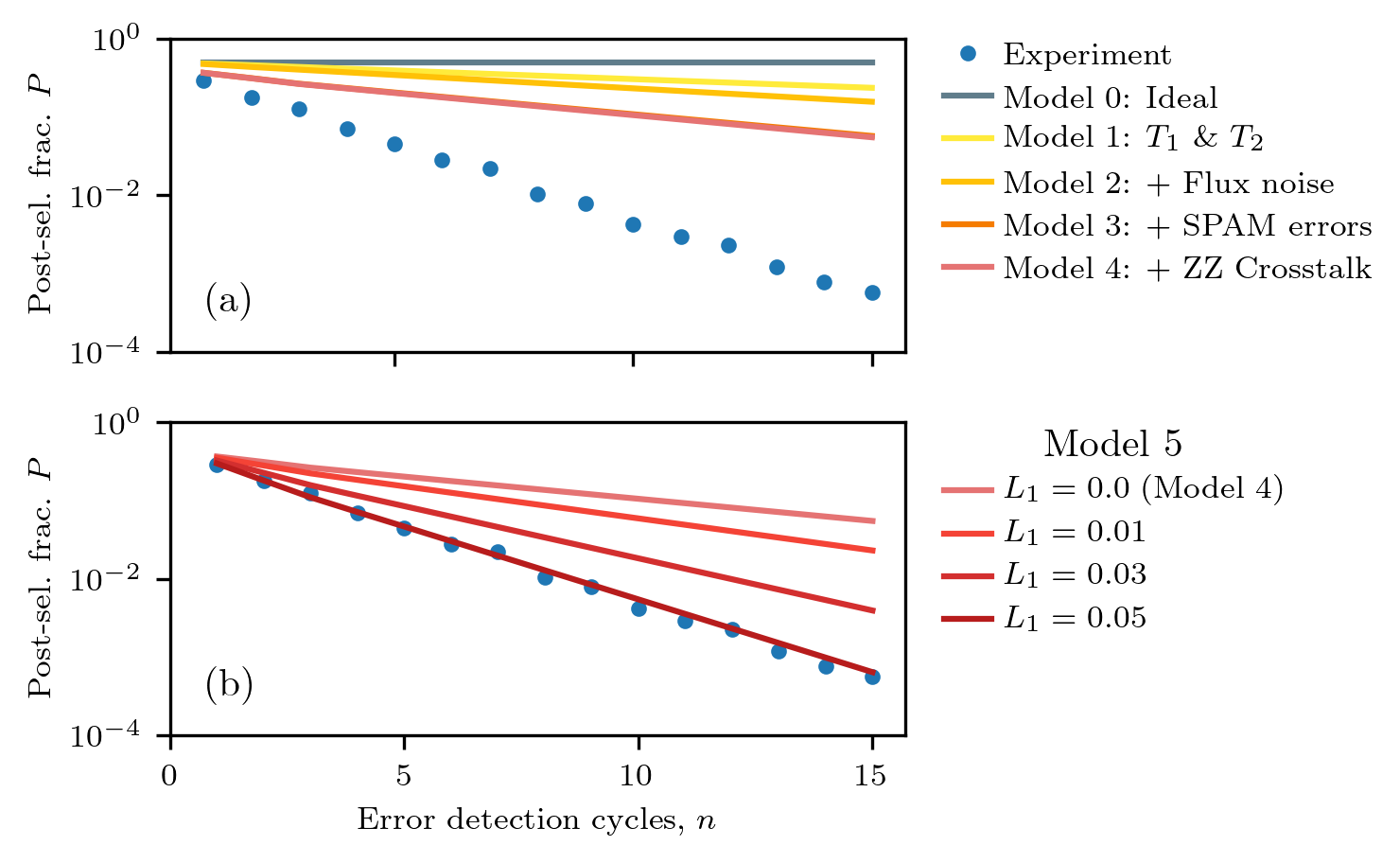}
    \caption{Post-selected fraction $P$ as a function of the number $n$ of error-detection cycles. The experimental $P$ (blue dots) is compared to numerical simulation under various models (solid curves).
         (a) Simulated $P$ obtained by incremental addition of error sources starting from the no-error (Model 0, gray); qubit relaxation and dephasing (Model 1, yellow); extra dephasing due to flux noise away from the sweetspot (Model 2, amber); state preparation and measurement errors (Model 3, orange); and crosstalk due to residual \(ZZ\) interactions (Model 4, red).
        (b) Simulated $P$ for Model 5 adding CZ gate leakage with 4 different values of $\leakrate$, the leakage per CZ gate, assumed equal for all CZ gates.}
    \label{fig:postsel_frac_model_comparison}
\end{figure}
We perform numerical density-matrix simulations using the~\textit{quantumsim} package~\cite{Obrien17} to study the impact of the expected error sources on the performance of the code.
We focus on repetitive error detection using the pipelined~scheme and with the logical qubit initialized in~$\zeroL$.
In~\cref{fig:postsel_frac_model_comparison}a, we show the post-selected fraction \(P(n)\) as a function of the number $n$ of error-detection cycles for a series of models.
Model 0 is a no-error model, which we take as the starting point of the comparison.
Model 1 adds amplitude and phase damping experienced by the transmon.
Model 2 adds  the increased dephasing away from the sweetspot arising from flux noise.
Model 3 adds  residual qubit excitation and readout (SPAM) errors.
Finally, Model 4 adds crosstalk due to the residual $ZZ$ coupling during the coherent operations of the stabilizer measurement circuits.
The details of each model and their input parameters drawn from experiment are detailed below.
We find that the dominant contributors to the error-detection rate are SPAM errors and decoherence.
However, we also observe that the noise sources included through Model 4 clearly fail to quantitatively capture the decay of the post-selected fraction observed in experiment.

We believe that an important factor behind the observed discrepancy is the presence of leakage, as suggested by the single-shot readout histograms in~\cref{fig:q_leak_histograms}.
We consider the leakage per CZ gate $\leakrate$ as a free parameter and assume the same value for all CZ gates.
We simulate the post-selected fraction for a range of $\leakrate$ values, shown in~\cref{fig:postsel_frac_model_comparison}b. We observe that $\leakrate\approx 5\%$ produces a good match with experiment, suggesting that leakage significantly impacts the error-detection rate observed. This value of $\leakrate$ is significantly higher than achieved in Ref.~\onlinecite{Negirneac20}, which used the same device. However, note that in this earlier experiment CZ gates were characterized while keeping all other qubits in $\ket{0}$. Spectator transmons with residual $ZZ$ coupling to either of the transmons involved in a CZ gate can increase $\leakrate$ when not in $\ket{0}$ (which is certainly the case in the present experiment). Note that leakage may also be further induced by the measurement~\cite{Sank16}, an effect that we do not consider in our simulation. However, the assumption that all CZ gates have the same $\leakrate$, the approximations used in our models, and other error sources that we have not considered here may lead to an overestimation of the true $\leakrate$.

Leakage is an important error source to consider in quantum error correction experiments of larger distance codes, requiring either post-selection based on detection~\cite{Varbanov20} or the use of leakage reduction units~\cite{Battistel21}.
We leave the detailed investigation of the exact leakage rates in our experiment and the mechanisms leading to them to future work.

\subsection{Error models}
Lastly, we detail the error models used in the numerical simulations in~\cref{fig:postsel_frac_model_comparison}.

\subsubsection{Model 1}
We take into account transmons decoherence by including an amplitude-damping channel parameterized by the relaxation time \(\Tone\) and a phase-damping channel parameterized by the pure-dephasing time at the sweetspot
\begin{equation*}
    \frac{1}{\Tphi^{\mathrm{max}}} = \frac{1}{\Ttwoecho} - \frac{1}{2\Tone},
\end{equation*}
where \(\Ttwoecho\) is the echo dephasing time (see~\cref{tab:Device}).
The qutrit Kraus operators defining these channels are detailed in Ref.~\onlinecite{Varbanov20} and we similarly introduce these channels during idling periods and symmetrically around each single-qubit or two-qubit gate (each period lasting half the duration of the gate).

\subsubsection{Model 2}
We consider the pure-dephasing rate~\(1/\Tphi = 2\pi\sqrt{\ln 2A}D_{\phi} + 1/\Tphi^{\mathrm{max}}\) away from the sweetspot due to the fast-frequency components of the $1/f$ flux noise, where $D_{\phi}$ is the flux sensitivity at a given qubit frequency and $A$ is the scaling parameter for the flux-noise spectral density. We use a $\sqrt{A}\approx 3~\mu\Phi_{0}$, the average of the extracted $\sqrt{A}$ values for $\Dthree$, $\Zone$ and $\X$ obtained by fitting the measured decrease of $\Ttwoecho$ as a function of the applied flux bias, following the model described above.
This allows us to estimate the dephasing time at the CZ interaction and parking frequencies, which then parameterize the applied amplitude-phase damping channel inserted during those operations~\cite{Varbanov20}.
We neglect the slow-frequency components of the flux noise due to the use of sudden Net Zero pulses, which echo out this noise to first order~\cite{Rol19,Negirneac20}.

\subsubsection{Model 3}
We further include state-preparation and measurement errors.
We consider residual qubit excitations, where instead of the transmon being initialized in $\ket{0}$ at the start of the experiment it is instead excited to $\ket{1}$ with a probability $p_{\mathrm{e}}$.
We extract $p_{\mathrm{e}}$ for each transmon from a double-Gaussian fit to the histogram of the single-shot readout voltages with the transmon nominally initialized in $\ket{0}$~\cite{Walter17}.
We model measurement errors via the POVM operators $M_{i} = \sum_{j=0}^{2}\sqrt{\prob\left(i\vert j\right)}\ket{j}\bra{j}$ for $i \in {0, 1, 2}$ being the measurement outcome, while $\prob\left(i\vert j\right)$ is the probability of measuring the qubit in state $\ket{i}$ when having prepared state $\ket{j}$.
We extract the probability $\prob\left(\qubit = \ket{i}\right)=\mathrm{Tr}\left(M_{i}^{\dagger}M_{i}\rho\right)$ of measuring qubit $\qubit$ in state $\ket{i}$ from simulation, where $\rho$ is the density matrix, while application of the POVM transforms $\rho \rightarrow M_{i}\rho M_{i}^{\dagger}/\prob\left(\qubit = \ket{i}\right)$.
In our simulations we condition on the detection of no error and thus we calculate $\prob\left(\qubit = \ket{0}\right)$ and then apply $M_{0}$ to the state $\rho$.
We obtain $\prob\left(0\vert j\right)$ for $j \in {0, 1}$ from the experimental assignment fidelity matrix~\cite{Heinsoo18} (where a heralded initialization protocol was used to prepare the qubits in $\ket{0}$~\cite{Riste12}) and we assume $\prob\left(0\vert 2\right) = 0$, consistent with the observed histograms in~\cref{fig:q_leak_histograms}.
At the end of each experiment with $n$ error-detection cycles we calculate the probability $\prob_{n}^{f}$ of obtaining trivial syndromes from the final measurements of the data qubits (see Results).
From this and from the probability $\prob_{n}\left(\anc_{i} = \ket{0}\right)$ of measuring ancilla $\anc_{i}$ in $\ket{0}$ at cycle $n$, we calculate the post-selected fraction of experiments defined as $P\left(n\right)= \prob_{n}^{f}\prod_{n}\prod_{i=1}^{3}\prob_{n}\left(A_{i} = \ket{0}\right)$.

\subsubsection{Model 4}
We consider the crosstalk due to residual $ZZ$ interactions between transmons.
The CZ gates involved in a parity check are jointly calibrated to minimize phase errors for the whole check as one block (see~\cref{fig:parity_checks}).
Instead of modeling this crosstalk as an always-on interaction and taking into account the details of the check calibration, we instead capture the net effect of this noise by including it as single-qubit and two-qubit phase errors in each CZ gate.
This assumes that the crosstalk only occurs between transmons that are directly coupled, which the measured frequency shifts observed in~\cref{fig:zz_matrix} validate.
We characterize the phases picked up during the CZ gates using $k\times 2^{k-1}$ Ramsey experiments for a check involving a total of $k$ transmons (including the ancilla).
In each experiment, we perform a Ramsey experiment on one transmon labelled $\qubit_{k}$. $\qubit_{k}$ is initialized in a maximal superposition using a $R^{-\pi/2}_{x}$ pulse, while the remaining $k-1$ transmons are prepared in each of the $2^{k-1}$ computational states $\ket{l}$.
Following this initialization, the parity check is performed, followed by a rotation of $R^{-\pi/2}_{\phi}$ (while the other transmons are rotated back to $\ket{0}$) and by a measurement of $\qubit_{k}$.
By varying the axis of rotation $\phi$, we extract the phase $\phi^{k}_{\mathrm{Ram}}\left(l\right)$ picked up by $\qubit_{k}$ with the remaining transmons in state \(\ket{l}\).
We perform this procedure for each of the $k$ transmons of the check, resulting in a total of $k\times 2^{k-1}$ measured phases, which are arranged in a column vector $\phivec^{\mathrm{Ram}}$.
We parameterize each CZ gate used in the parity check by a matrix $\mathrm{diag}\left(1, e^{i\phi_{01}},e^{i\phi_{10}}, e^{i\phi_{11}}\right)$.
The column vector $\phivec^{\mathrm{CZ}}$ then contains all of the phases parameterizing each of the $k-1$ CZ gates involved in the parity checks, with $k=3$ for the \(\ZD{1}\ZD{3}\) and \(\ZD{2}\ZD{4}\) checks and $k=5$ for the $\ZD{1}\ZD{2}\ZD{3}\ZD{4}$ check.
We can express each of the measured phases in the Ramsey experiment as a linear combination of the acquired phases as a result of the CZ interactions between transmons, i.e., $\phivec^{\mathrm{Ram}} = A\phivec^{\mathrm{CZ}}$, where the matrix $A$ encodes the linear dependence.
Given the measured $\phivec^{\mathrm{Ram}}$ we perform an optimization to find the closest $\phivec^{\mathrm{CZ}}$ as given by
\begin{subequations}
    \begin{align*}
         & \underset{\phivec^{\mathrm{CZ}}}{\text{min}} & \sum_{i}\left(\sum_{j}A_{ij}\phivec^{\mathrm{CZ}}_{j} - \phivec^{\mathrm{Ram}}_{i}\right)^{2} , \\
         & \text{subject to}                            & 0 \leq  \phivec^{\mathrm{CZ}}_{j} < 2\pi.
    \end{align*}
\end{subequations}
The optimal $\phivec^{\mathrm{CZ}}$ then captures the net effect of the $ZZ$ crosstalk during the parity checks, which we include in the simulation.
We do not model phase errors accrued during the ancilla readout, since in our simulation we condition on each ancilla being measured in $\ket{0}$.

\subsubsection{Model 5}
We model leakage due to CZ gates following the model and numerical implementation presented in Ref.~\onlinecite{Varbanov20}.
Here, we do not consider the phases picked up when non-leaked transmons interact with leaked ones (the leakage-conditional phases~\cite{Varbanov20}) and we set them to their ideal values.
We also neglect higher-order leakage effects, such as excitation to higher-excited states or leakage mobility.
Thus, we only consider the exchange of population between $\ket{11}$ and $\ket{02}$ given by $4\leakrate$, except for the CZ between $\Zone$ and $\Dthree$, where the population is instead exchanged with $\ket{20}$ as we use the $\ket{11}$-$\ket{20}$ avoided crossing for this gate~\cite{Negirneac20}.

There remain several relevant error sources beyond those included in our numerical simulation.
For example, we do not include dephasing of data or other ancilla qubits induced by ancilla measurement, which we expect to be a relevant error source for comparing the performance of the pipelined and parallel schemes.
Also, we only consider the net effect of crosstalk due to residual $ZZ$ interactions during coherent operations of the parity-check circuits, which we include via errors in the single-qubit and two-qubit phases in the CZ gates.
Thus, we do not capture the crosstalk present whenever an ancilla is projected to state $\ket{1}$ by the readout but declared to be in $\ket{0}$ instead.
Furthermore, as $ZZ$ crosstalk does not commute with the amplitude damping included during the execution of the circuit, we are not capturing the increased phase error rate that this leads to.

%
\end{bibunit}

\end{document}